\documentclass[journal]{IEEEtran}
\usepackage{graphicx}
\usepackage{cite} 
\usepackage{subcaption}
\usepackage{xcolor}
\usepackage{lipsum}
\usepackage{multicol}
\usepackage{amsmath}
\usepackage{amssymb}
\usepackage[utf8]{inputenc}
\usepackage{caption}
\usepackage[labelsep=period]{caption}
\usepackage[font=small,labelfont=small]{caption}
\makeatletter
\newcommand*{\rom}[1]{\expandafter\@slowromancap\romannumeral #1@}
\makeatother

\ifCLASSINFOpdf
 
\else
  
\fi

\begin{document}
\title{	 A Wireless and Battery-free  Biosensor Based on Parallel Resonators for  Monitoring a  Wide Range of Biosignals}

\author{Rehab S. Hassan,~\IEEEmembership{Member,~IEEE}

\thanks{R. S. Hassan is with  the Department of Electrical and Electronic
	Engineering, Yonsei University, Seoul 120749, South Korea  e-mail: (rehab@yonsei.ac.kr).}}

\maketitle

\begin{abstract}
	
This paper proposes a novel wireless, battery-free, and label-free biosensor
	for minimally invasive and non-invasive permittivity sensing for applications such as detecting glucose levels in the interstitial dermal fluid. The miniaturized, fully passive sensor is based on two symmetric parallel 0.8 mm$^3$ LC (inductor-capacitor) resonators. Each inductor is integrated with a conductive plate at one of its terminals. The passive sensor's main field is dominant outside the resonators, creating an effective capacitance outside the sensor, unlike reported in previous studies where the effective capacitance was limited to between the resonator lines.  The proposed sensor was further analyzed under two different scenarios to change the region of effective capacitance and test its suitability for  pressure monitoring, such as wound monitoring; first, by repositioning one of the two resonators, and second, by replacing one of the resonators with a conductive plate.  The experimental results confirmed the proposed passive sensor's performance in detecting the glucose concentration in an aqueous solution with a sensitivity of  500 and 46 kHz/(mg/dL)  for minimal invasive and non-invasive monitoring, respectively, within the glucose range of 0-500 mg/dL with volume sample requirements as low as 60 $\mu$L. Further, by repositioning one of the resonators, the effective capacitance lies inside the passive sensor, making it suitable for wound monitoring. The results indicated that the resonance shift can be made fairly linear with respect to the variation in the separation between the resonators with a sensitivity of 2.5$\times 10^3$ MHz/mm within the separation range  of  0.2-0.5 mm, considering separation as the main factor to sense the pressure or healing of the wound.

\end{abstract}

\IEEEpeerreviewmaketitle

\section{Introduction}
\IEEEPARstart{M}{icrowave} 	 spectroscopy is a powerful  tool for monitoring anomalies in  biological tissues that can detect a type of cancer, burn depth, and blood sugar levels \cite{mirbeik2017ultra, khokhar2017near, adhikari2015ultrahigh, hussein2019breast, topfer2015millimeter, shah2019improved, matthieu2014evolution, yilmaz2019radio}. Microwave spectroscopy is label-free, non-intrusive,  antibody-free, and uses non ionizing radiation, which  makes it highly appropriate for biomedical applications. Biological tissues, like any other material, exhibit  unique dielectric properties and electric polarization (P), described as permittivity, which  determines the degree of interaction  of electrical polarization of a material with the applied external electric field  \cite{chen2004microwave}. Any variation  in the permittivity of biological tissues  mainly occurs due to a change in  its electric polarization, therefore indicating an anomaly. Thus,  microwave spectroscopy is one of the promising diagnosis  techniques.
 
\begin{figure}[t]
	\centering
	\includegraphics[width=3.5in]{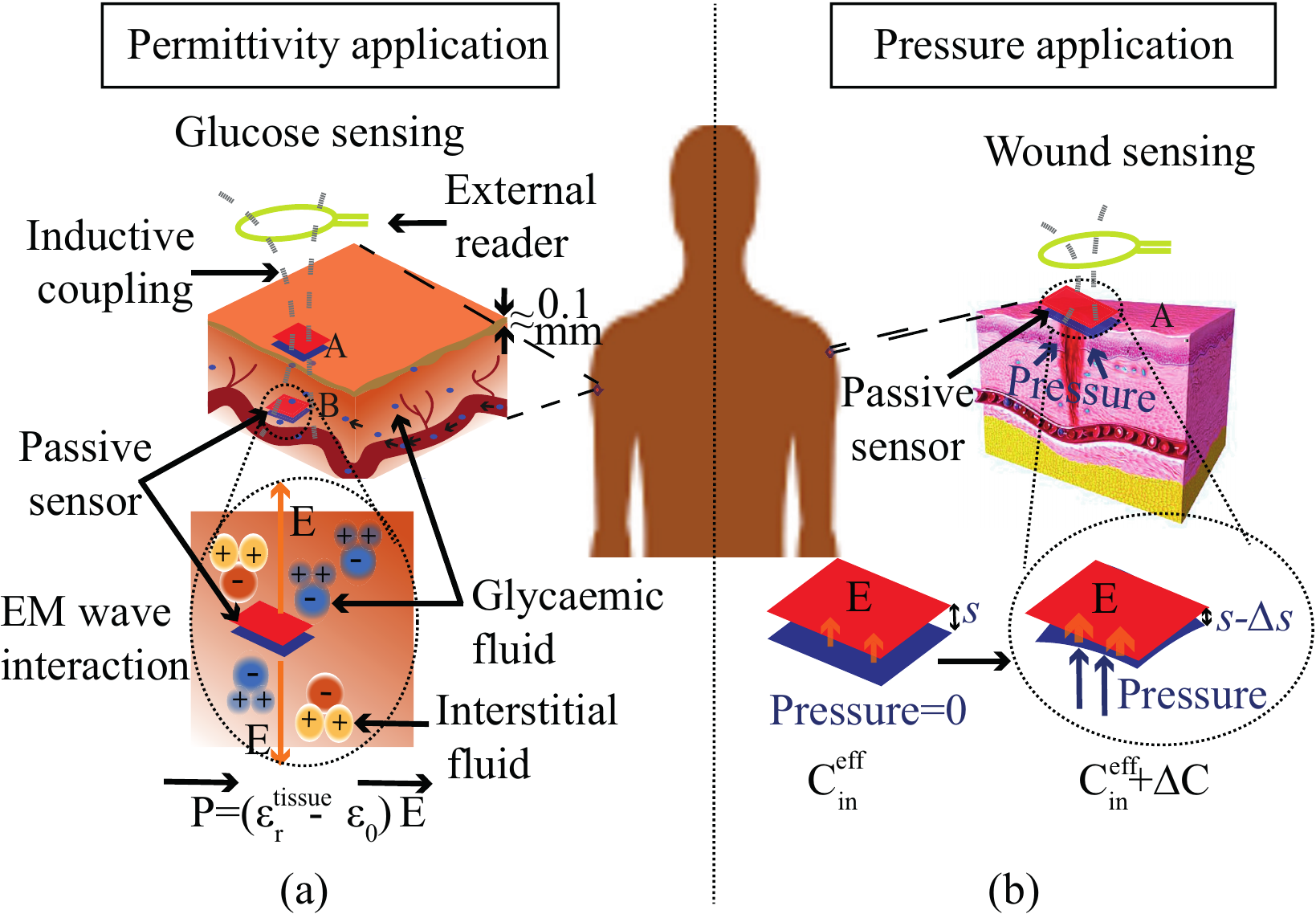}
	\caption{ Schematic  of the fully passive resonator-based biosensor for wearable applications  ( i.e., position A) or minimally invasive  ( i.e., position B) (a) Glucose monitoring  (b) Wound monitoring}
	\label{fig1}
\end{figure}

In general, microwave spectroscopy techniques can be classified into resonant and non-resonant techniques  for the characterization of  biological tissues. The resonator technique is most commonly used in biomedical applications because it accurately indicates  changes observed in biological tissues. The resonator-based biosensor is based on the interaction of the biosignals with either the main or  fringe field. Although the fringe field can be used to monitor changes in the  biosignals \cite{caduff2006non, caduff2003first, pandit2021towards}, it has a low sensitivity in comparison with that of  the main field; and hence, the main field is preferred  for monitoring of biosignals \cite{ huang2018microstrip}.  Recently, many \textit{wired} \textit{active} resonator-based biosensors have been designed to confine the electric fields between resonator lines in a  specific region with  high potential difference, by using devices, such as an interdigital capacitor  {\cite{kim2015rapid, kim2019inter,  kazemi2020ultra} or a split ring resonator \cite{ lee2008biosensing, kandwal2020highly,   omer2020low, omer2020non},  provided the biological tissues are  dropped in that specific region. Thus,  to facilitate the use of these biosensors in  wearable applications, the biosignals  interact with the  fringe fields, while the main field is located between the resonator lines.  As a result, the sensitivity of the biosensors reduces because the biosignals are not located between the resonator lines. \textit{Thus, a paradigm shift is required to acquire a strong electric field outside the resonator without  locating biosignals between the resonator lines.}   The above issue is addressed in this study by designing  a novel biosensor based on two symmetric parallel LC resonators, where the strength of the    electric field is dominant outside the biosensor, unlike in the previous studies  in which the main field was confined between the resonator  lines  \cite{caduff2006non, caduff2003first, pandit2021towards, huang2018microstrip, kim2015rapid, kim2019inter, kazemi2020ultra,   lee2008biosensing, kandwal2020highly, omer2020low, omer2020non, govind2020design, kiani2021dual}. Additionally, we used a \textit{wireless}, \textit{battery-free}  resonator-based biosensor to monitor the biological tissues owing to its compact size, light weight, cost efficient fabrication, and convenient use in  wearable and implantable applications \cite{huang2016lc,  yeon2019optimal}.}


Over the last decade, wireless,  battery-free sensing systems have gained significant attention in the biomedical field. The system comprises a  passive resonator that interacts with biological tissues,  and an external reader that tracks the resonance frequency of the passive resonator outside the tissues. Thus, many studies have focused on improving either the passive resonator or the external reader. The application of microelectromechanical system (MEMS) technologies has been widely introduced for improving the structure and performance of passive resonators. This system monitors the continuous wireless physiological parameters for various types of pressure by using the passive  resonator \cite{huang2019emerging}, and focuses on using passive resonator comprising of advanced materials such as air cavities \cite{chen2010wireless, tan2017wireless,  yeh2019fabrication}, flexible pyramidal dielectric materials \cite{deng2018lc,deng2019symmetric,chen2014continuous }, or flexible sponge materials \cite{kou2019wireless} to improve the performance of the resonator. The passive resonator-based pressure monitoring  used in these studies are based on the parallel-plate technique, with either an inductor on the top of the plate, \cite{chen2010wireless, tan2017wireless,  yeh2019fabrication, deng2018lc,deng2019symmetric}, or  using two parallel inductors \cite{chen2014continuous, kou2019wireless}.  Furthermore, based on wireless power transfer technologies, improving the wireless power transfer to the passive resonator improves the ability of the external reader to track the resonant frequency of the passive resonator. For example, replacing a vector
network analyzer with a portable reader has been introduced \cite{wang2017novel}. Moreover, systems, such as the parity-time (PT) symmetry and exceptional points   (EPs), have also been studied for improving wireless power transfer to the passive resonator-based biosensors to enhance the sensing capability of the biological tissues \cite{chen2018generalized, dong2019sensitive}. However, none of these studies have explained the reason and the advantages behind using the specified resonator structures owing to shortcomings in the analysis and equivalent circuit.



This paper raises  a key question: Does any passive sensor based on parallel plates, where the plates are   separated by a distance $s$,  induce an effective capacitance  inside the sensor? Considering this,   the proposed passive sensor was further analyzed by changing one of the resonators, either by repositioning it or replacing it with a conductive plate to serve as three different passive sensors. Three passive  sensors based on the parallel plate technique were studied. The direction of current through the inductor  and  the charge distribution were taken into consideration  to investigate the direction of  uniform electric field between the plates, and effective capacitance.  Consequently,  the sensing performance of the sensor was improved for both pressure sensing (i.e., wound monitoring) and
permittivity sensing (i.e., glucose monitoring).  To the best of our  knowledge, \textit{this paper is the first to analyze the region of dominance of the  electric field in various passive sensors}. Therefore,  the key contributions of this study include

 \begin{itemize}
 	  \item A  conceptual understanding of  the region of dominance of the  electric field in   different passive sensors to monitor wide-range biosignals is presented. As this  topic is largely ignored,  we believe that it could be beneficial in designing passive resonator-based biosensors.

  \item A novel approach to analyze the glucose level in the fluid  is developed. It can operate underneath the skin for a minimal invasive or  non-invasive approach, as shown in Fig. \ref{fig1}(a).  Based on the two parallel resonating plates, the uniform electric field (main field) outside the passive sensor is considered to be  dominant and thus, the capacitance outside the sensor  is effective.

\end{itemize}
   
 \begin{figure}[!b]
 	\centering
 	\includegraphics[width=3.4in]{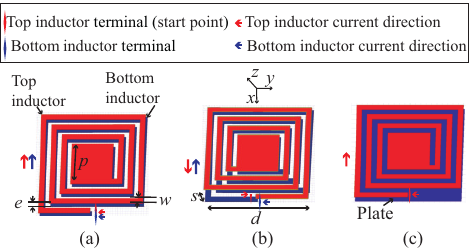}
 	\caption{  Top view of  the passive  sensor  (a) proposed sensor: two parallel inductors turn clockwise with the same start point and current direction flow through both inductors (sensor 1). (b) Repositioning one of the inductors of proposed sensor: top inductor turns  counter clockwise while the bottom inductor turns  clockwise with opposite start point and, current direction flow through the inductors is opposite to each other (sensor 2).  (c)  Replacing  one of the inductors of the proposed sensor with a conductive plate (sensor 3).}
 	\label{fig2}
 	\vspace{5pt}
 \end{figure}


\section{System analysis and  principle of operation }
The proposed passive sensor is based on two symmetric parallel  inductors, as shown in Fig. \ref{fig2}(a). Further, a change in any one of the  inductors, such as repositioning or being replaced with a conductive plate,  is studied, as shown in Fig. \ref{fig2}(b) and (c). Three  miniaturized passive sensors were designed, simulated, and fabricated on a  2 $\times$ 2 $\times$ 0.2 mm$^3$ plane flame retardant 4 (FR4) substrate.  An FR4 has a relative permittivity of 4.4 and a loss tangent of 0.02. The passive sensors were designed using a 3D full-wave finite element simulator (ANSYS HFSS) and fabricated  using standard printed circuit board technology.   A single turn loop coil was used as external reader,  which was placed outside the biological tissues, as shown in Fig. \ref{fig1}(a) and (b). Based on the near field (non-radiative technique) the  passive sensors  were inductively powered using the external reader.

\begin{figure}[h]
	\centering
	\includegraphics[width=3in]{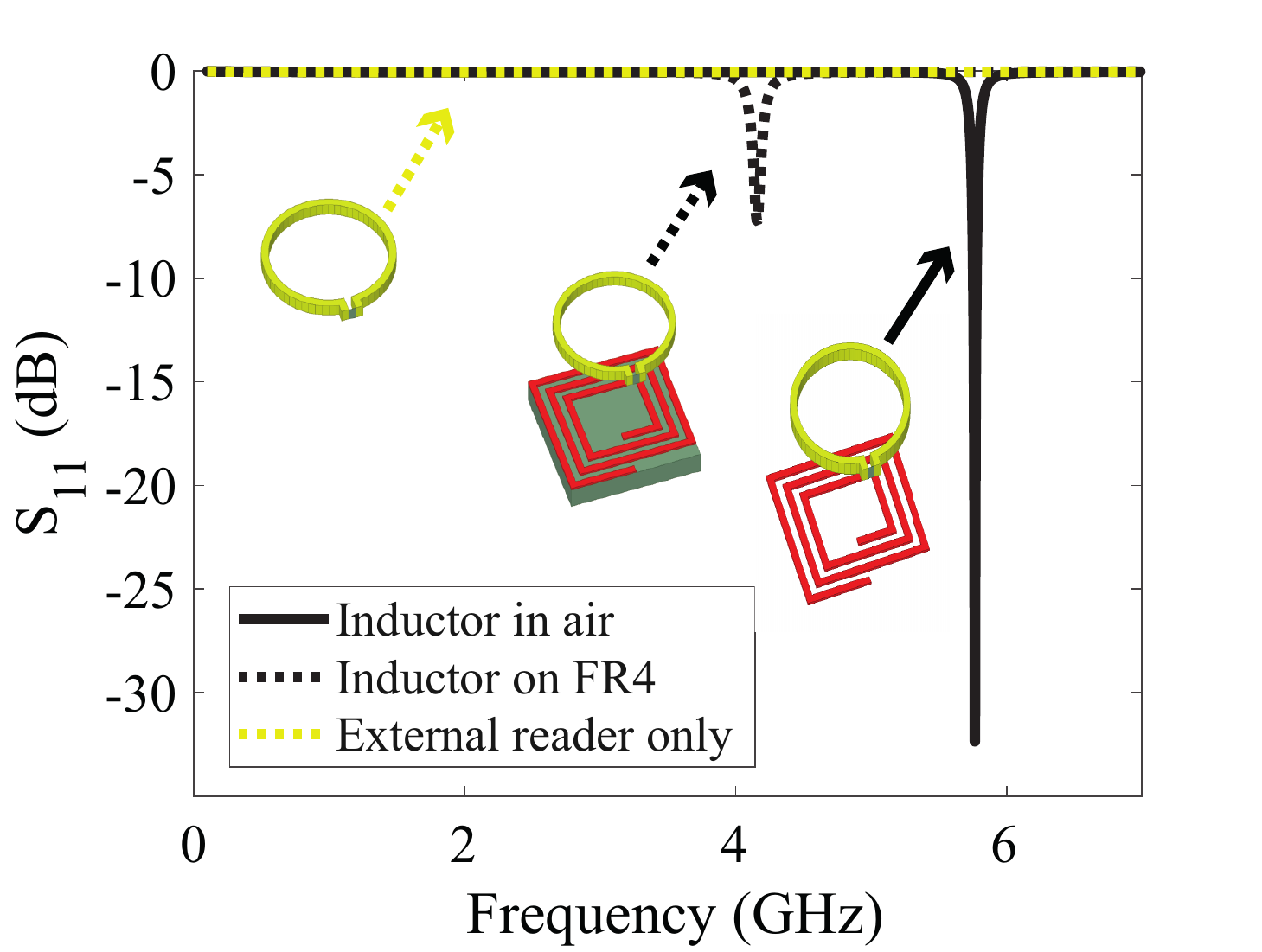}
	\caption{ HFSS-simulated resonant frequency of the spiral  inductor of the passive sensor at a distance of 1 mm from the external reader}
	\label{inductor}
\end{figure}

   Table \ref{table1} lists the parameters of the sensor's inductor.  The top  and bottom inductor  are symmetric; and hence,   $L_1$  = $L_2$ = $L_s$, where $L_s$ represents the inductance of  the passive sensor, and  can be defined by:
\begin{equation}
	L_{s} = \frac{2\mu_0}{\pi}n^2d_{avg}(\ln(\frac{2.07}{\phi})+ 0.18\phi+0.13\phi^2)
	\label{ls}
\end{equation}
where $d_{avg}=\frac{(d+d_{in})}{2}$ is  the average diameter of the square, and $\phi=\frac{1-\alpha}{1+\alpha } $. $\alpha$  is the ratio of the outer diameter to the inner diameter \cite{mohan1999simple, lee2003design}. $\mu_0$  is the permeability of the free space ($\mu_0 =4\pi\times10^{-7}$~N.A$^{-2}$). The  inductance of  the passive sensor  L$_s$ was approximately calculated as  20 nH. Furthermore, the  distributed capacitance of a spiral inductor was  studied in \cite{wu2003analysis,  huang2006scalable}. Generally for any pattern, the distributed capacitance is given as:
\begin{equation}
	\frac{1}{C_0} = \frac{1}{4\epsilon_0Q^2}\int\int \left[\frac{\rho(r_i)\rho(r_j)K_{ij}}{\big|r_i-r_j\big|}\right]d^3r_id^3r_j
	\label{c0}
\end{equation}
The integration is applied over the entire area of the pattern. $Q$ is a total charge, and  $\rho(r_i)$ is the spatial charge density as a function of $r_i$, which is the length of the inductor trace. $\epsilon_0$  is the permittivity of the free space ($\epsilon_0 =8.85\times10^{-12}$~F.m$^{-1}$).  The distributed capacitance C$_0$ of the inductor of the passive sensor was approximately calculated  as 0.05 pF~\cite{woodard2010functional, kurs2007wireless, mohammed2019noninvasive}. Thus, from (\ref{ls}) and (\ref{c0}),  the self-resonance frequency of the spiral inductor of the proposed sensor was calculated, where the resonance for the inductor in the free space was given as $f_0=1/2\pi\sqrt{C_0L_s}$.  Table \ref{table22} shows a comparison of  the calculated resonant frequency and the simulation results.  Fig. \ref{inductor} shows the resonant frequency of the spiral inductor of the passive sensor through the external reader.  Further, the frequency decreases to 4.2 GHz when the inductor is positioned at the top of FR4, considering a part of the fringe capacitance of the inductor lines interacts with FR4  \cite{wu2003analysis, huang2006scalable, kurs2007wireless, woodard2010functional, mohammed2019noninvasive, jow2009modeling}.

Furthermore, the capacitive coupling and resonance of the passive resonators are studied based on the parallel plate technique. Additionally, the direction of current through the inductor depending on the external reader coil is analyzed to understand the charge distribution on the inductor, considering it could improve the sensitivity of the sensor towards the pressure and permittivity of applications.

\begin{table}[t]
	\caption{ Geometrical parameters of the  inductor }
	\centering
	\vspace{5pt}
	\begin{tabular}{c  c c } 
		\hline	\hline
		Parameter &	Description & Values (unit) \\ [0.5ex] 
		\hline
		$n$ & Number of turns  & 3    \\[1ex]
		$e$ & Line spacing     & 0.1 mm   \\[1ex]
		$w$ & Line width     & 0.1 mm   \\[1ex]
		$t$ & Line thickness     & 0.017 mm   \\[1ex]
		$p$ & Inner plate   & 0.8 mm \\[1ex]
			d & Outer diameter   & 2 mm \\[1ex]
		\hline	\hline
	\end{tabular}
	\label{table1}
	\vspace{-5pt}
\end{table}

\begin{table}[t]
	
	\caption{  Characteristic frequencies of a planar spiral inductor }
	\centering
	\vspace{5pt}
	\begin{tabular}{c  c c c } 
		\hline	\hline
		Frequency &Calculated in air & Simulated in air & Simulated on FR4\\ [0.5ex] 
		\hline
		Values (GHz) &$f_0$=5.1  & $f_{0cal}$=5.7&$f_{sub}$=4.2    \\[1ex]
		\hline 	\hline
	\end{tabular}
	\label{table22}
\end{table}
\begin{figure*}[ht]
	\centering{\includegraphics[width=180mm]{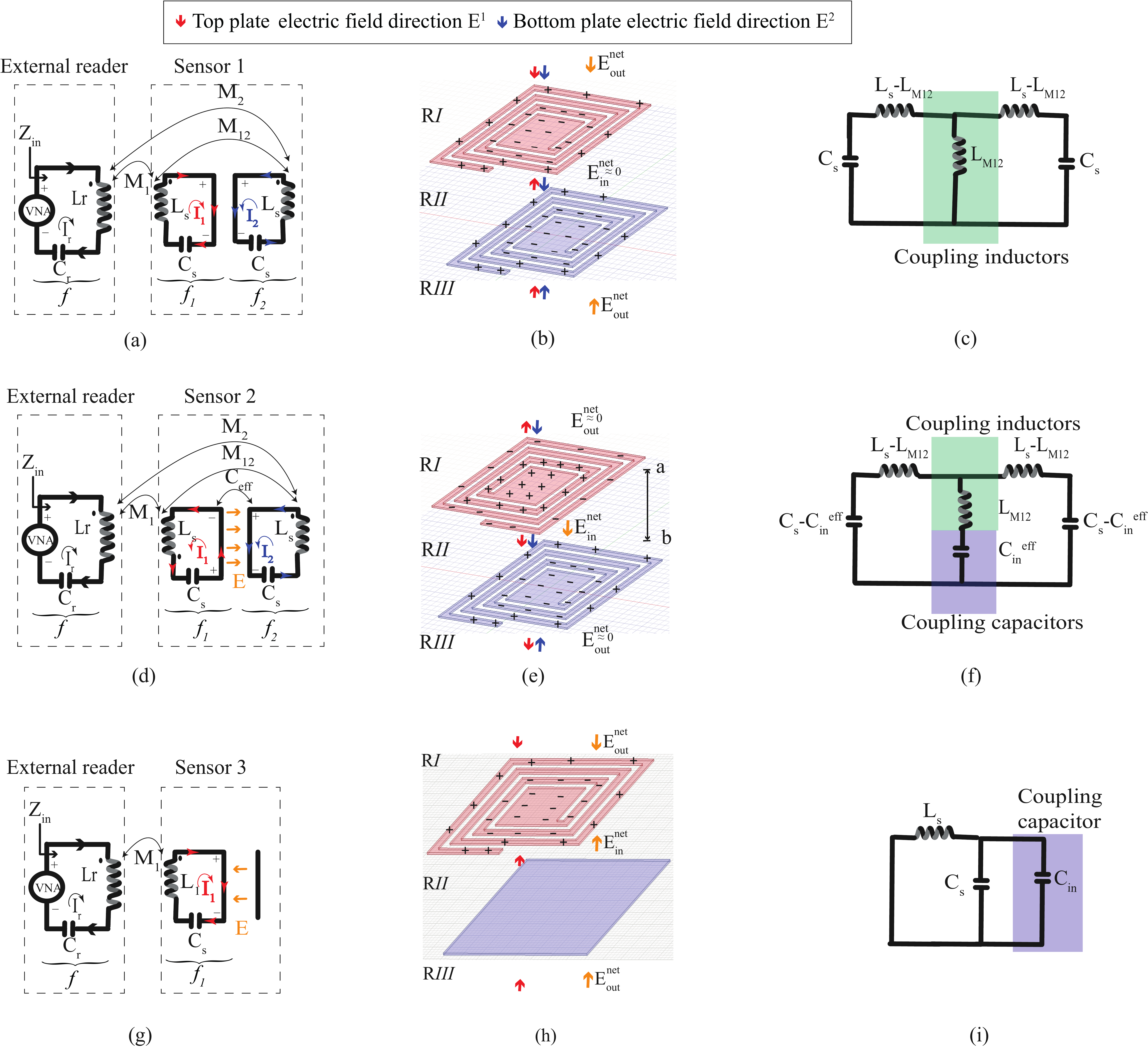}}
	\caption{Circuit analysis of different passive sensors inductively coupled through an external reader inductor  (a) Sensor 1 (d) Sensor 2 (g) Sensor 3.  Charge distribution of the inductor,  and the  electric field of the plate based on Gauss's law~\cite{woodard2010functional, kurs2007wireless, duran2012electrically, feynman1975lectures}~(b) Sensor 1 (e) Sensor 2 (h) Sensor 3.    Schematic diagram of the miniaturized passive sensor (c) Sensor 1 (f) Sensor 2 (i) Sensor 3 }\label{circuit}
\end{figure*}
\subsection{Two-symmetric inductor with same current direction}
\begin{figure}[b]
	\centering{\includegraphics[width=85mm]{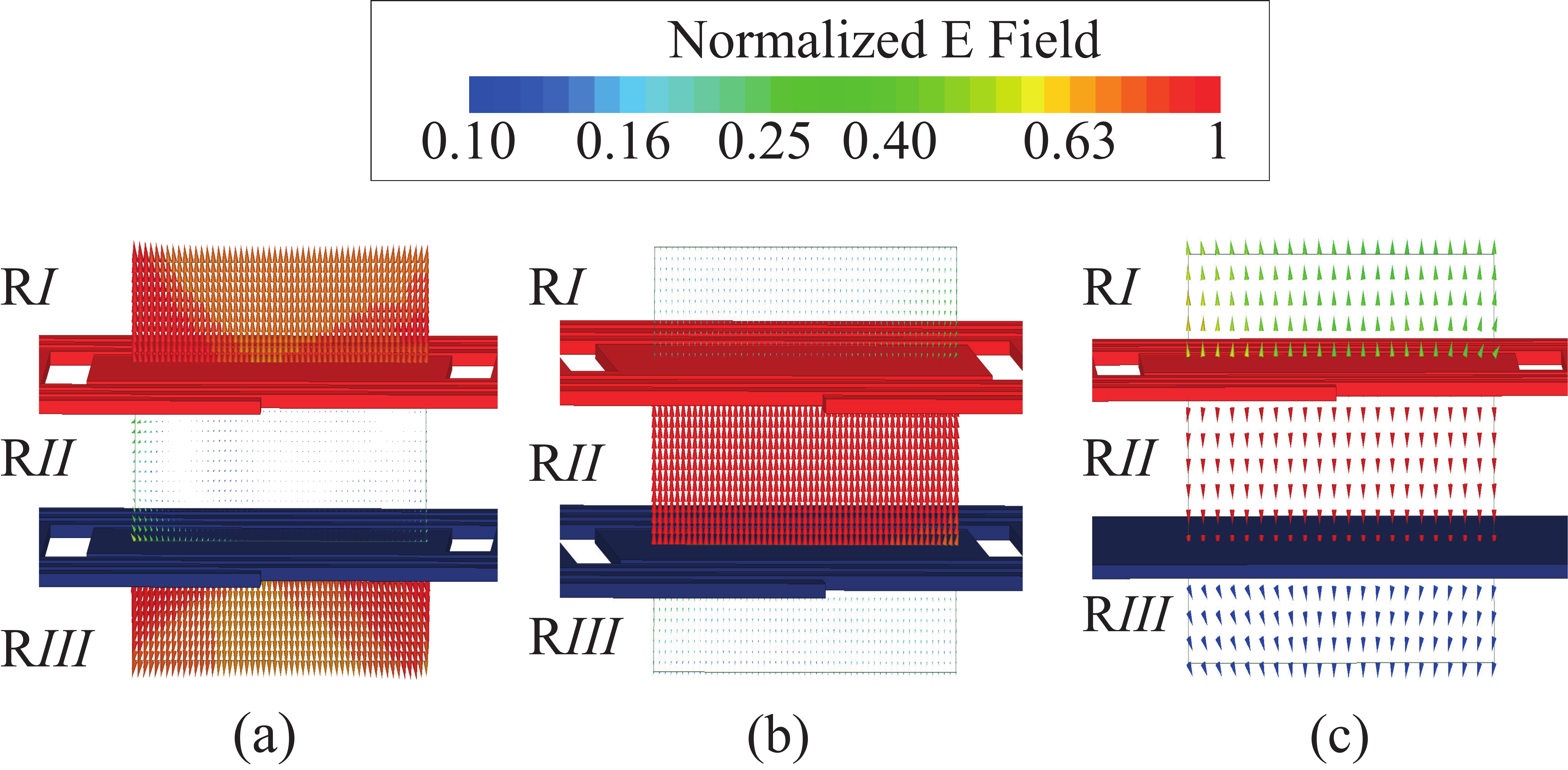}}
	\caption{ HFSS-simulated  normalized electric field distribution between the inner plates of the inductors of the passive sensors (a) Sensor 1 (b) Sensor 2 (c) Sensor 3}\label{field}
\end{figure}

Based on the parallel plate technique,  the two symmetric spiral inductors of the proposed passive  sensor  have the same turn direction (clockwise), as shown in  Fig. \ref{fig2}(a).  Additionally, the  current in both inductors also  move in the same direction, therefore  creating identically  charged inductors.  Thus, according to Gauss’s law, there is not enough  potential difference between the two inductors, and the direction of  the electric field is reversed to the outside the sensor.  As a result, there is no effective capacitance between the two-symmetric inductors, which means no energy is stored inside the sensor. Fig. \ref{circuit}(a)  illustrates  the circuit analysis by applying the  dot convention, which  represents the turn direction  of the inductors, which are adjacent to each other.  The current leaving the dotted terminal of the  top inductor  is in-phase with the current leaving the dotted terminal of the  bottom inductor, and hence the polarities of the voltages at both dotted  terminals   are also in-phase. Therefore, when the voltage is positive at the dotted terminal of the top inductor, the voltage at the dotted terminal of the  bottom inductor  is also positive. Fig. \ref{circuit}(b) demonstrates the uniform electric field between the inner surface  of the  inductors plates at region II (R\emph{II}), which according to the principle of superposition given as:
\begin{equation}
\oint \vec E_{in}\cdot \vec da=E_{in}^{net}a 
\end{equation}
\begin{equation}
{E}_{in}^{net}={E}_{in}^1+{E}_{in}^2=-\frac{Q_1/2}{\epsilon a}  +\frac{Q_2/2}{\epsilon a} \approx 0
	\label{sensoe1_Ein}
\end{equation}
where a and $\epsilon$ represent the surface area of the plate and the  permittivity at R\emph{II}, respectively, and the net electric field between the induction lines of both inductors is found to be zero. However, the net uniform field outside the passive sensor (at the top and bottom sides) is  dominant.  The  electric field of the inner surface  of the  inductors  plates at the top,  at region I (R\emph{I}) can be represented as:
\begin{equation}
{E}_{out}^{net}={E}_{out}^1+{E}_{out}^2=\frac{Q_1/2}{\epsilon_0 a}  +\frac{Q_2/2}{\epsilon_0 a} = \frac{Q}{\epsilon_0 a}
	\label{sensoe1_Eout1}
\end{equation}
where  the permittivity of free space $\epsilon_0$  is outside the passive sensor. Similarly,  the electric  field at the bottom side,  at region III (R\emph{III}) is given as:
\begin{equation}
{E}_{out}^{net}=- \frac{Q}{\epsilon_0 a}.
\label{sensoe1_Eout3}
\end{equation}
From the above analysis, it is observed that  electric field is dominant on the  outside of the proposed sensor (sensor 1), as seen in Fig. \ref{field}(a). Thus,  it is useful for permittivity  applications where the  biosignal interacts directly with the sensor. The main field is used to form an effective  capacitance with the biological tissues, where the outer capacitance $C_p$ represents the ability of biological tissues to store energy received  from an external electric field.  A change in the permittivity of the biosignal  $\Delta \epsilon_r^{tissue}$, causes the  outside capacitance to  change  $\Delta C_{p}$, $ C_{p}\approx  (\epsilon_r^{tissue})C_{p}^{air}$.


Using Kirchhoff's voltage law (KVL ),   the input impedance $Z_{in}$ of the sensor 1 is derived  from the external reader (as seen in Fig. \ref{circuit}(a)), which is given as:
\begin{gather}
	\begin{bmatrix} V \\ 0 \\ 0 \end{bmatrix}
	=j
	\begin{bmatrix}
		\omega L_r-\frac{1}{\omega C_r}  &-\omega  L_{M_1} & \omega L_{M_2} \\
		-\omega L_{ M_1}  & \omega L_s-\frac{1}{\omega C_s} & -\omega L_{M_{12}}  \\
		-\omega L_{M_2}  &-\omega L_{M_{12}} & \omega L_s-\frac{1}{\omega C_s}  
	\end{bmatrix}
	\begin{bmatrix} I_r \\ I_1 \\ I_2 \end{bmatrix}
\end{gather}
\begin{equation}
\begin{gathered}
Z_{in}(f) = \frac{V}{I_r}=j\bigg(\omega L_{r}-\frac{1}{\omega C_{r}} \bigg)-\bigg(\frac{\omega ^2 L_{M_2}^2}{j(\omega L_{s}-\frac{1}{\omega C_{s}})} \bigg)
+\\\bigg(\frac{-\omega L_{M_1}(\omega L_{s}-\frac{1}{\omega C_{s}} )-\omega^2 L_{M_{12}}L_{M_{2}}}{\omega^2 L_{M_{12}}^2-(\omega L_{s}-\frac{1}{\omega C_{s}})^2}\bigg)\\
\times \bigg(-\frac{\omega^2 L_{M_1}L_{M_2}}{j(\omega L_{s}-\frac{1}{\omega C_{s}} )}-j\omega L_{M_1}\bigg)
\end{gathered}
\label{z1}
\end{equation}
 When the external reader is  coupled to the passive sensor,  then $f$ $\approx$$f_s$, where the second, third, and fourth terms on the right-hand side in (\ref{z1}) causes  a peak rise in the real part of the input impedance.  Also, the reflection coefficient is represented as: 
\begin{equation}
S_{11} =20~log\mid \frac{Z_{in} -Z_{0} }{Z_{in} + Z_{0} }\mid
\end{equation}
where $Z_{0}$,  $L_{r}$, and $C_{r}$  are the characteristic impedance,  inductance, and  capacitance of the external reader, respectively \cite{pozar2009microwave}. $L_{M_{1}}$, $L_{M_{2}}$, and $L_{M_{12}}$  are the mutual inductance between the external reader and the top inductor, the external reader and the bottom inductor, and the top and bottom inductors of the sensor, respectively.

 $C_s$ represents the equivalent capacitance of the passive sensor, which in turn represents $C_0$, fringe capacitance, and the capacitance $C_p$  of the outer surface of the inductor plate.   From equations  (\ref{sensoe1_Eout1}) and (\ref{sensoe1_Eout3}), we see that the electric field of the outer surface of  the inductors plates is dominant, creating an effective capacitance,  $C_s$   is dominant.   From equation (\ref{sensoe1_Ein}), we see that the capacitance between the inductors  plates is negligible because the electric field between the sensor is quite weak.  The sensor resonates  when the reactance  is zero ($Img (Z_s)=0$). Fig. \ref{circuit}(c) shows the sensor circuit analysis whose  resonant frequency can be expressed as:
\begin{equation}
	f_{s} \approx \frac{1}{2\pi \sqrt{(L_s \pm L_{M_{12}}) C_s}}
	\label{h}
\end{equation}
In loose coupling between the external reader coil and the proposed passive  sensor in the near field, the  frequency splitting can not be observed. Thus, the resonance frequency  can be  further expressed as:
\begin{equation}
	f_{s} \approx \frac{1}{2\pi \sqrt{(L_s + C_s)}}.
	\label{h}
\end{equation}
Considering  the direction of the electric field of the inductor  plate is  outside the sensor  towards infinity,   $C_p$ will not affect the distributed capacitance, and hence, $f_s$=$f_{sub}$=$f_1$=$f_2$. Further, by introducing the biological tissue, $C_p$ is effective. In turn, $C_s$ is effective, and the resonance frequency of the proposed sensor  giving as:
\begin{equation}
	f_{s}(\epsilon_r^{tissue}) \approx \frac{1}{2\pi \sqrt{(L_s + C_s(\epsilon_r^{tissue}))}}.
	\label{h}
\end{equation}
\subsection{Two-symmetric inductor with opposite current direction}
Herein, repositioning one of the inductor of the proposed passive sensor is studied (sensor 2), as shown in Fig. \ref{fig2}(b). The turn direction of the  top and  bottom inductors is counter-clockwise and  clockwise, respectively. Therefore, the  current direction of the inductors is opposite to each other, which creates a high potential difference and  an effective  capacitance  $C_{in}^{eff}$ lies inside the inductors. Thus, the inductors  store charges in between them and  are electrically separated. The circuit analysis of sensor 2 is shown in Fig. \ref{circuit}(d) where the position of the dots on the terminal of each inductor is different, that is, at the opposite ends of the turn direction which indicates  that the top and bottom inductors turns  in opposite directions. As a result, the current leaving  the dotted terminal of  top inductor is 180$^o$ out-of-phase with the current  leaving the dotted terminal of   the bottom inductor, indicating that  the polarities of the voltages at the dotted terminals  are also out-of-phase.

 Therefore, when the voltage is positive at the dotted terminal of the bottom inductor, the voltage across the  top inductor will be negative. This induces a high capacitance between the two inductors, where the net electric field on the inner surface of the  inductors plates  at R\emph{II}, as shown in Fig. \ref{circuit}(e), is given as:
\begin{equation}
	\oint \vec E_{in} \cdot \vec da=E_{in}^{net}a 
\end{equation}

\begin{equation}
	{E}_{in}^{net}={E}_{in}^1+{E}_{in}^2=\frac{Q_1/2}{\epsilon a} +\frac{Q_2/2}{\epsilon a} = \frac{Q}{\epsilon}
	\label{sensoe2_E}
\end{equation}
 Fig. \ref{field}(b) shows the electric field on  the inner surface of the inductors plates. The electric potential is given as: 
\begin{equation}
	V_{in}=E\int_{a}^{b}~dL=Es=\frac{Qs}{\epsilon}
\end{equation}
The capacitance  between the two inductors plates at  R\emph{II} can be expressed as:
\begin{equation}
	C_{in}^{eff(p)} \approx \frac{Q}{V} \approx\frac{\epsilon a}{s}
	\label{sensoe2_C}
	\vspace{-5pt}
\end{equation}
Further,  the overall inner capacitance between the inductors  is given as:
\begin{equation}
	C_{in}^{eff} \approx\frac{\epsilon A}{s}
	\label{sensoe2_Call}
	\vspace{-5pt}
\end{equation}
where A represents the surface area of spiral metal trace and plate.  Changes in the electrical separation $s$ at any position of the inductor by an outside pressure or force lead to a linear change in the capacitance corresponding to the self-resonance frequency of sensor 2.  Further,  as observed in Fig. \ref{circuit}(d), the input impedance $Z_{in}$ is given as:
\begin{gather}
	\begin{bmatrix} V \\ 0 \\ 0 \end{bmatrix}
	=j\\
	\begin{bmatrix}
		\omega L_r-\frac{1}{\omega C_r}   & \omega M_1 & \omega M2 \\
		\omega M_1  & \omega L_s- \frac{1}{\omega C_s}- \frac{1}{\omega C_{in}^{eff}}  & \omega  M_{12}+\frac{1}{\omega C_{in}^{eff}} \\
		-\omega M_2  & \omega M_{12}+\frac{1}{\omega C_{in}^{eff}} & \omega L_2-\frac{1}{\omega C_2}-\frac{1}{\omega C_{in}^{eff}} 
	\end{bmatrix}\\
	\begin{bmatrix} I_r \\ I_1 \\ I_2 \end{bmatrix}
\end{gather}

\begin{equation}
	\begin{gathered}
		Z_{in}(f) = j\bigg(\omega L_{r}-\frac{1}{\omega C_{r}} \bigg)-\bigg(\frac{\omega ^2 M_2^2}{j(\omega L_{s}-\frac{1}{\omega C_{s}} -\frac{1}{\omega C_{in}^{eff}})}\bigg)+ \\
		\bigg(\frac{\omega M_1(\omega L_{s}-\frac{1}{\omega C_{s} }-\frac{1}{\omega C_{in}^{eff}} )+ \omega M_2(\omega M_{12}+\frac{1}{ C_{in}^{eff}})}{
			-(\omega L_{s}-\frac{1}{\omega C_{s}}-\frac{1}{\omega C_{in}^{eff}})^2+(\omega M_{12}+\frac{1}{ C_{in}^{eff}})^2}\bigg)\\
		\times \bigg(\frac{\omega M_2(\omega M_{12}+\frac{1}{\omega C_{in}^{eff}} )}{j(\omega L_{s}-\frac{1}{\omega C_{s}} -\frac{1}{\omega C_{in}^{eff}} )}+j\omega M_1\bigg).
		\label{z}
	\end{gathered}
\end{equation}
Similarly, from the sensor circuit analysis shown in Fig. \ref{circuit}(f),  the resonance frequency of  sensor 2  is  given as: 
\begin{equation}
	f_s\approx \frac{1}{2\pi \sqrt{L_s\pm L_{M_{12}} (C_s+ C_{in}^{eff}})} \approx \frac{1}{2\pi \sqrt{L_s (C_s+ C_{in}^{eff}})}.
	\label{sensoe2_f}
\end{equation}

\subsection{Inductor on the top of the plate}
This section discusses another modification that was made  to the proposed passive sensor, which is replacing one of the resonators with a plate (sensor 3), shown in Fig. \ref{fig2}(c).   The external reader is inductively coupled to the inductor with current flowing through the inductor,  as shown in Fig. \ref{circuit}(g). The electric field between the plates at R\emph{II}, as shown in Fig. \ref{circuit}(h) is given as:

\begin{equation}
{E}_{in}^{net}=- \frac{Q_1/2}{\epsilon a}
\end{equation}
The electric field outside the sensor at R\emph{I}  is given as:
\begin{equation}
{E}_{out}^{net}=- \frac{Q_1/2}{\epsilon_0 a}.
\end{equation}
As seen in Fig.  \ref{circuit}(h) and \ref{field}(c), the electric field is dominant partly outside and partly inside the sensor. Therefore, we could say that the induced capacitance  $C_{in}$ in the sensor was due to the electric field inside the sensor, which could be suitable for sensing pressure or permittivity. However, reduced sensitivity is expected because there is no specific region where the electric field is dominant. From the circuit analysis shown in Fig. \ref{circuit}(d), the input impedance can be expressed as:



\begin{gather}
	\begin{bmatrix} V \\ 0  \end{bmatrix}
	=j
	\begin{bmatrix}
	\omega L_r-\frac{1}{\omega C_r}  &  -\omega M_1 \\
		\omega L_s-\frac{1}{\omega C_s}-\frac{1}{\omega C_{in}} &  -\omega M_1 
	\end{bmatrix}
	\begin{bmatrix} I_r \\ I_1 \end{bmatrix}
\end{gather}

\begin{equation}
Z_{in}(f)=j(\omega L_r - \frac{1}{\omega C_r})+ (\frac{\omega^2 M_1^2}{j(\omega L_s - \frac{1}{\omega C_s}- \frac{1}{\omega C_{in}})})
\end{equation}
Also, Fig. \ref{circuit}(i) shows the circuit analysis of sensor 3, and the resonant frequency  is solved as:

\begin{equation}
Img (Z_s)=Img (Z_1)+Img (Z_2)=0
\end{equation}

\begin{equation}
0 = \bigg(\frac{1}{\omega C_s-\frac{1}{\omega L_s } } \bigg)+ \bigg(\frac{1}{\omega C_{in}}  \bigg)
\label{y}
\end{equation}
Thus, the resonant frequency is expressed as:
\begin{equation}
f_{s} \approx \frac{1}{2\pi \sqrt{L_s (C_s+C_{in}})}.
\label{66}
\end{equation}

\begin{figure}[!t]
	\centering{\includegraphics[width=70mm]{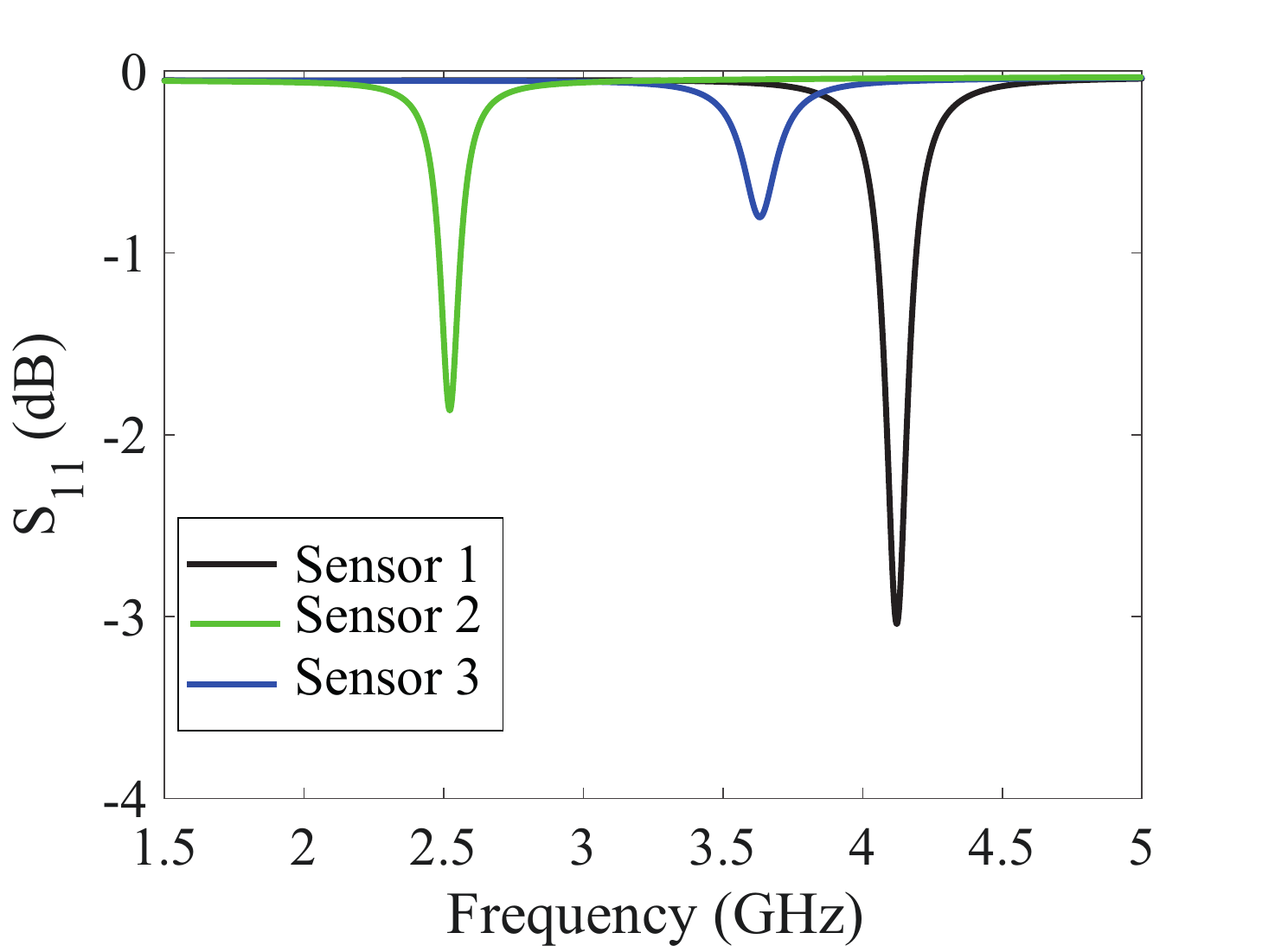}}
	\caption{ HFSS-simulated resonance frequency of  passive sensors }\label{resonance}
\end{figure}

\begin{figure}[h]
	\centering{\includegraphics[width=90mm]{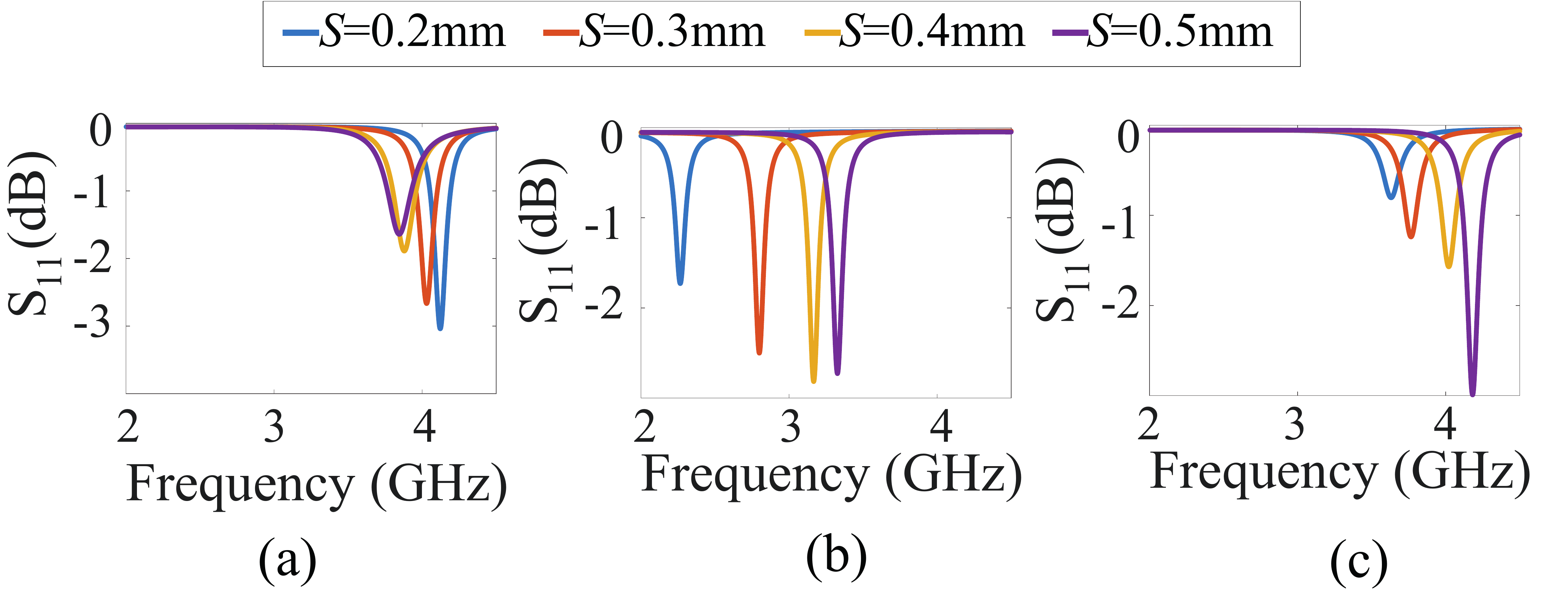}}
	\caption{  
		HFSS-simulated  resonance of the passive sensor by changing  separation $s$  between the plates by changing the dielectric material thickness.  (a) Sensor 1 (b) Sensor 2 (c) Sensor 3  }\label{all}
\end{figure}

\begin{figure}[h]
	\centering{\includegraphics[width=90mm]{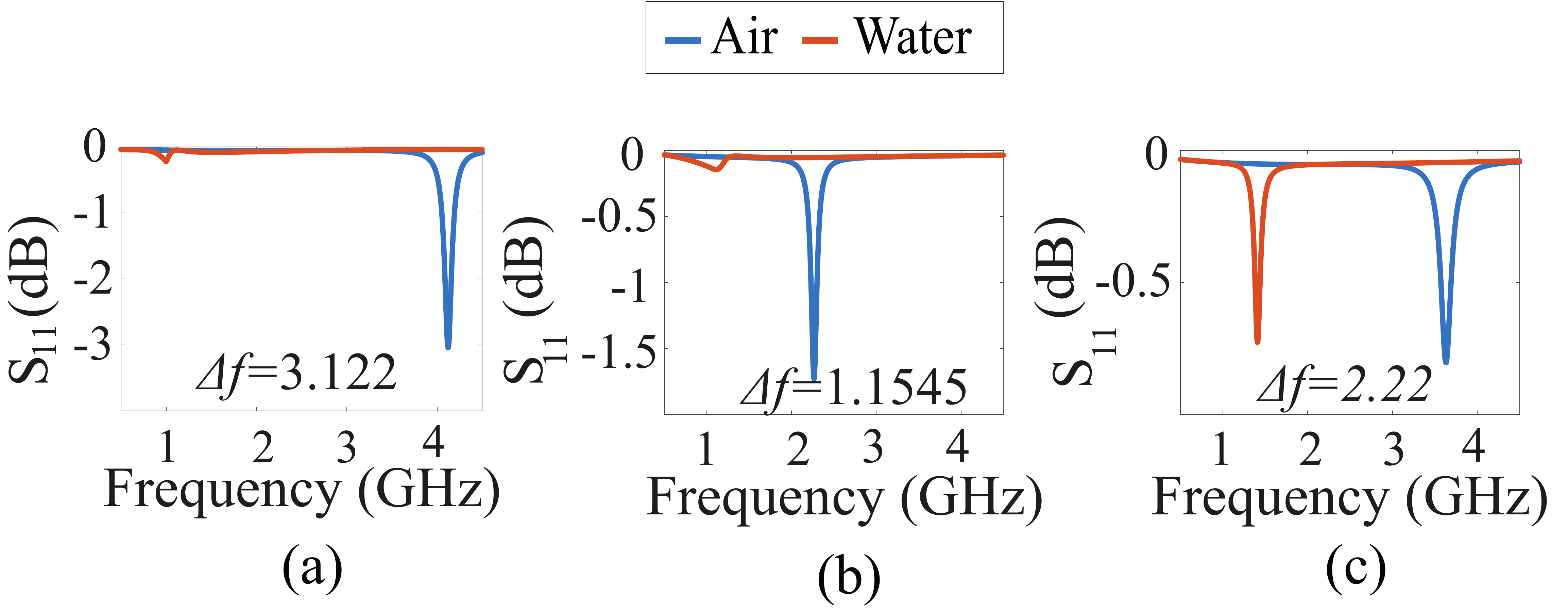}}
	\caption{  HFSS-simulated  resonance of the passive sensor with a thickness of 0.2 mm by changing the surrounding medium.  (a) Sensor 1 (b) Sensor 2 (c) Sensor 3}\label{per}
\end{figure}

Figure \ref{resonance} shows the self-resonance frequency of the three passive sensors with the presence of the FR4 slab between the plates. A large frequency shift is observed in sensor 2 as compared to sensors 1 and  3 due to the intensity  of the uniform electric field between the inductors, which produces $C_{in}^{eff}$ as given in  equation (\ref{sensoe2_Call}), thus increasing  the equivalent capacitance of sensor 2. 
However,  there is no frequency shift in the proposed passive sensor (sensor 1) because there is no effective capacitance between the inductors due to no electric field  (see Fig. \ref{field}(a)). Additionally, there is no energy stored in the substrate, which is an advantage in monitoring  the permittivity  change of the biosignals. Further,  the effect of change in the separation $s$ between plates is studied. Shown in Fig. \ref{all}(a),  no frequency shift  was observed for the proposed passive sensor (sensor 1) when  separation $s$ is reduced. However, the resonance frequency of sensor 2 decreases with decrease in the separation between the two plate, as seen in Fig. \ref{all}(b),  and proven in equations (\ref{sensoe2_C}) and (\ref{sensoe2_f}).  Although  the resonance frequency of sensor 3 decreases with change in the separation due to $C_{in}$, it has lower sensitivity than sensor 2, as shown in  Fig. \ref{all}(c). Furthermore, surrounding the sensors with lossy material such as water is studied. As shown in Fig. \ref{per}(a),  the proposed passive  sensor (sensor 1) shows the maximum resonance frequency shift due to the dominant  strength of the uniform electric field outside the sensor, which  creates an effective capacitance  with lossy material (water).  As a result, $C_s$ is effective. As shown in Fig.  \ref{per}(b), a slight resonance  shift is observed in  sensor 2  mainly due to the fringe field of the inductors lines. There is a noticeable resonance shift in sensor 3, as shown in Fig. \ref{per}(c).


\section{ Measurements and Results}
\subsection{ Pressure  monitoring (wound monitoring)}
\begin{figure}[!t]
	\centering{\includegraphics[width=50mm]{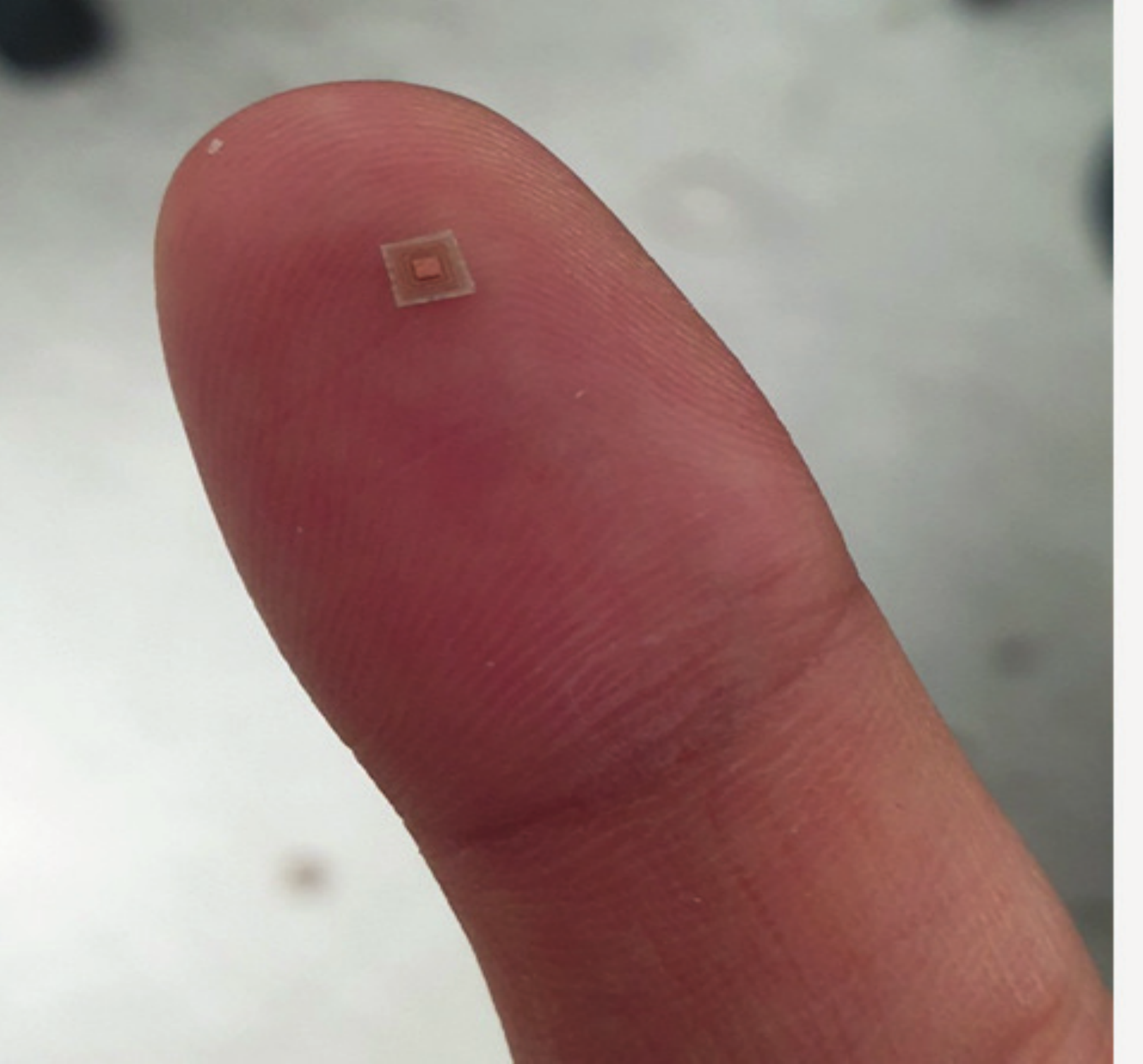}}
	\caption{ Top view of a  passive sensor}
	\label{setup}
\end{figure}

\begin{figure}[!t]
	\centering{\includegraphics[width=87 mm]{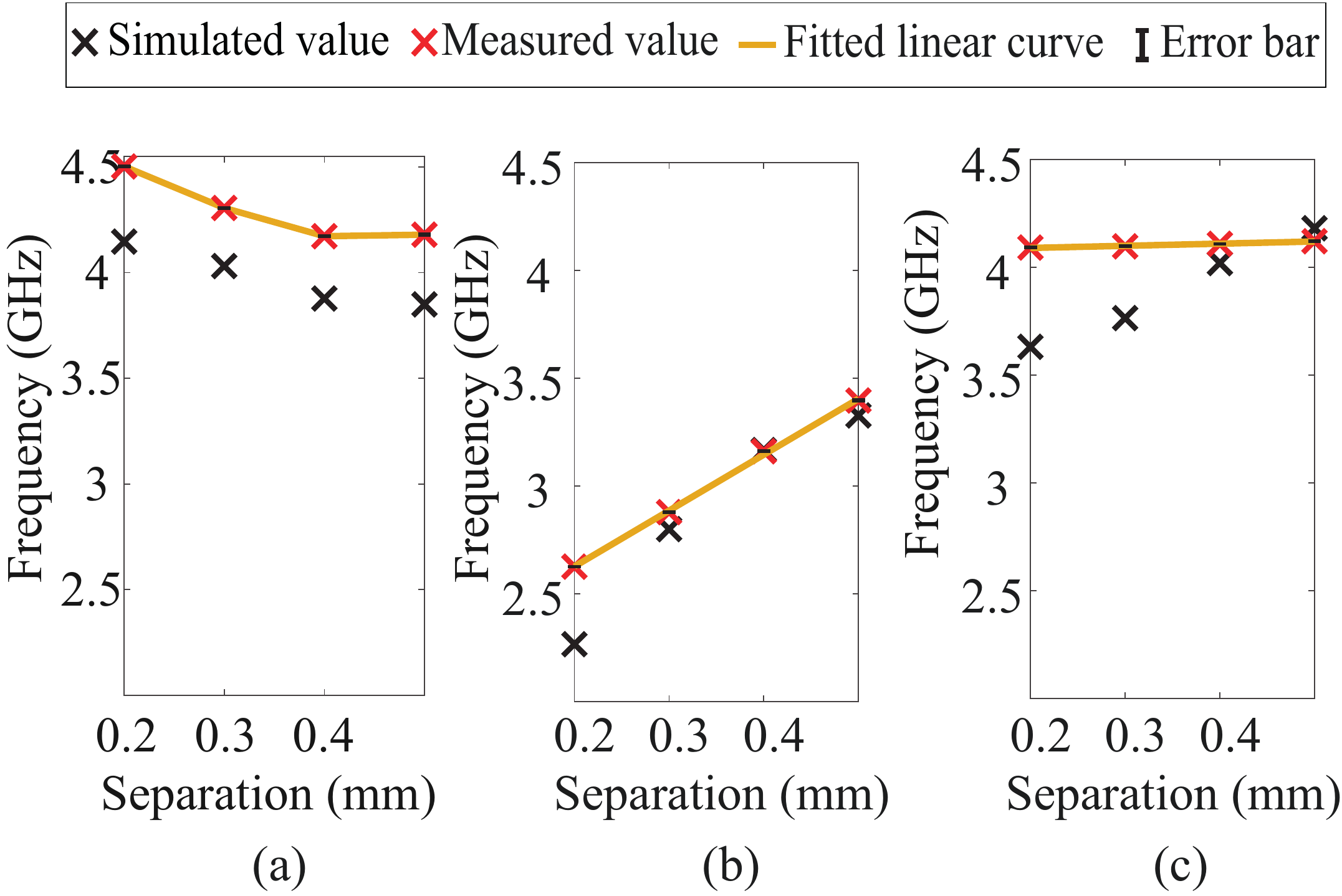}}
	\caption{  Measured and simulated the values of  the resonance frequency of passive sensor by changing the separation between the plates.  (a) Sensor 1 (b) Sensor 2 (c) Sensor 3 }\label{cap}
\end{figure}
\begin{figure}[!t]
	\centering{\includegraphics[width=60 mm]{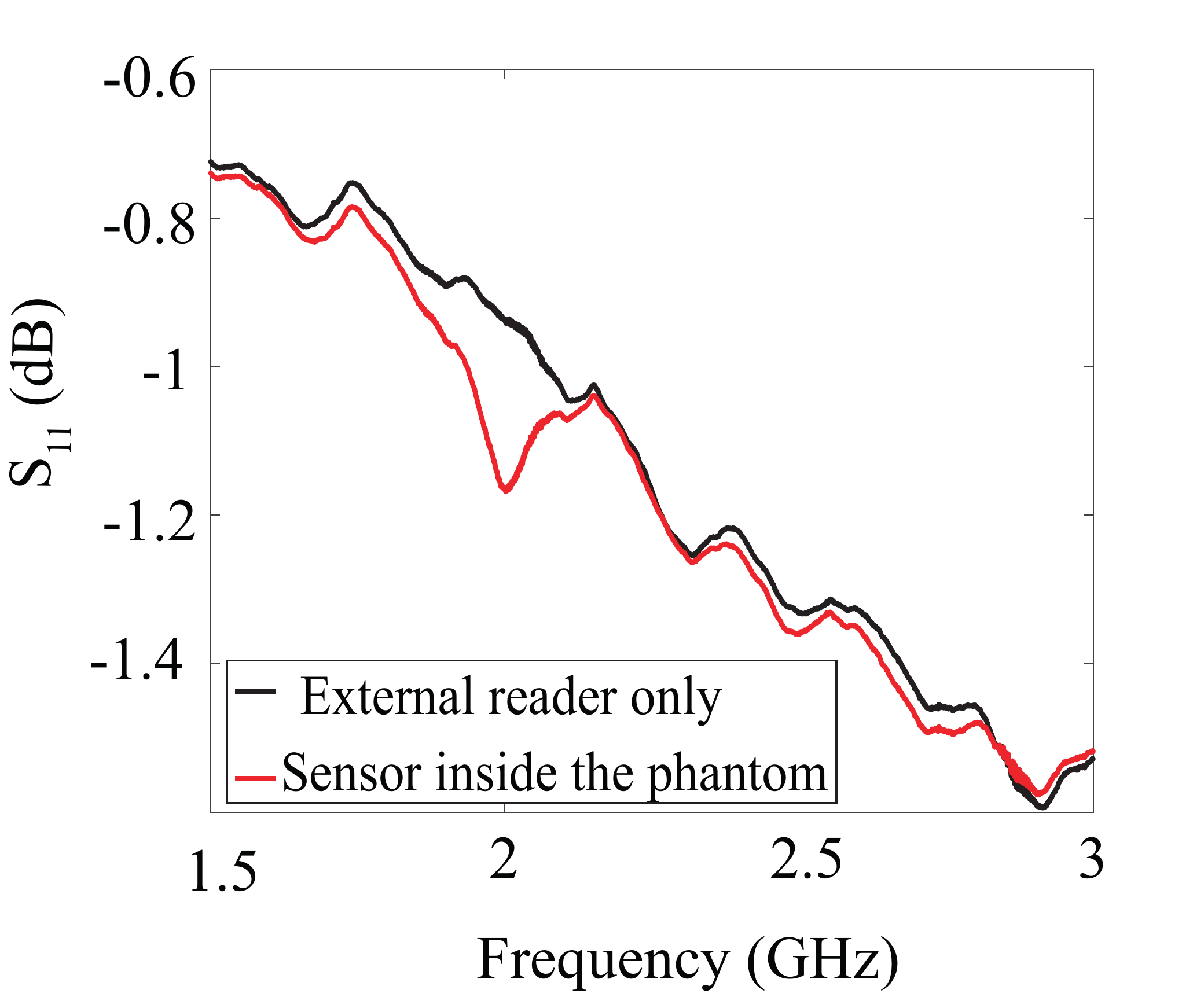}}
	\caption{External reader noise with sensor resonance frequency through phantom }\label{setup2}
\end{figure}
\begin{figure}[!t]
	\centering{\includegraphics[width=90mm]{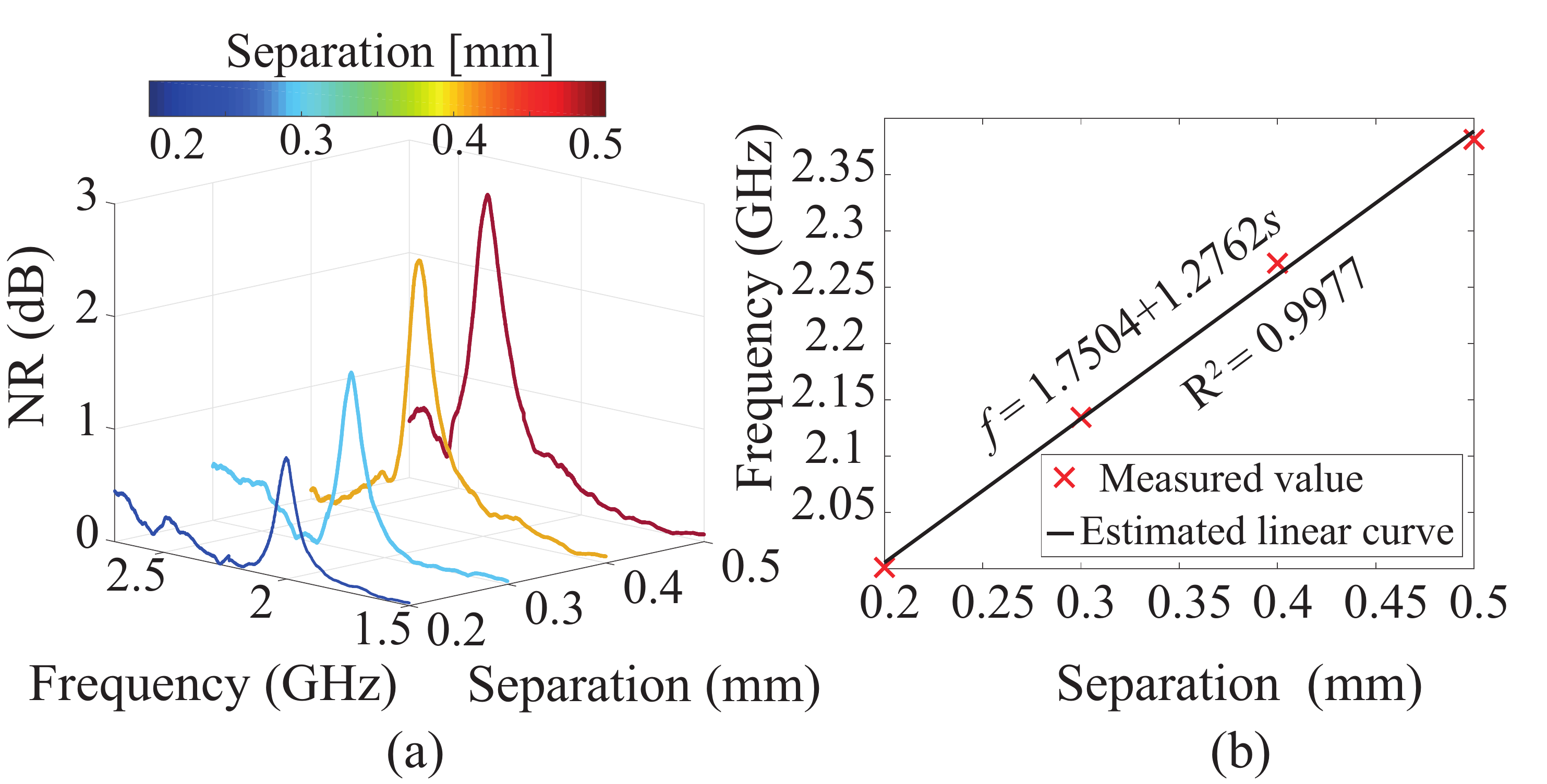}}
	\caption{  (a) Measured $S_{11}$ response of sensor 2  with different thicknesses inside the muscle tissue.  (b) Estimated linear values of resonant frequencies for different thicknesses.}\label{signals}
\end{figure}

\begin{figure}[!t]
	\centering{\includegraphics[width=90mm]{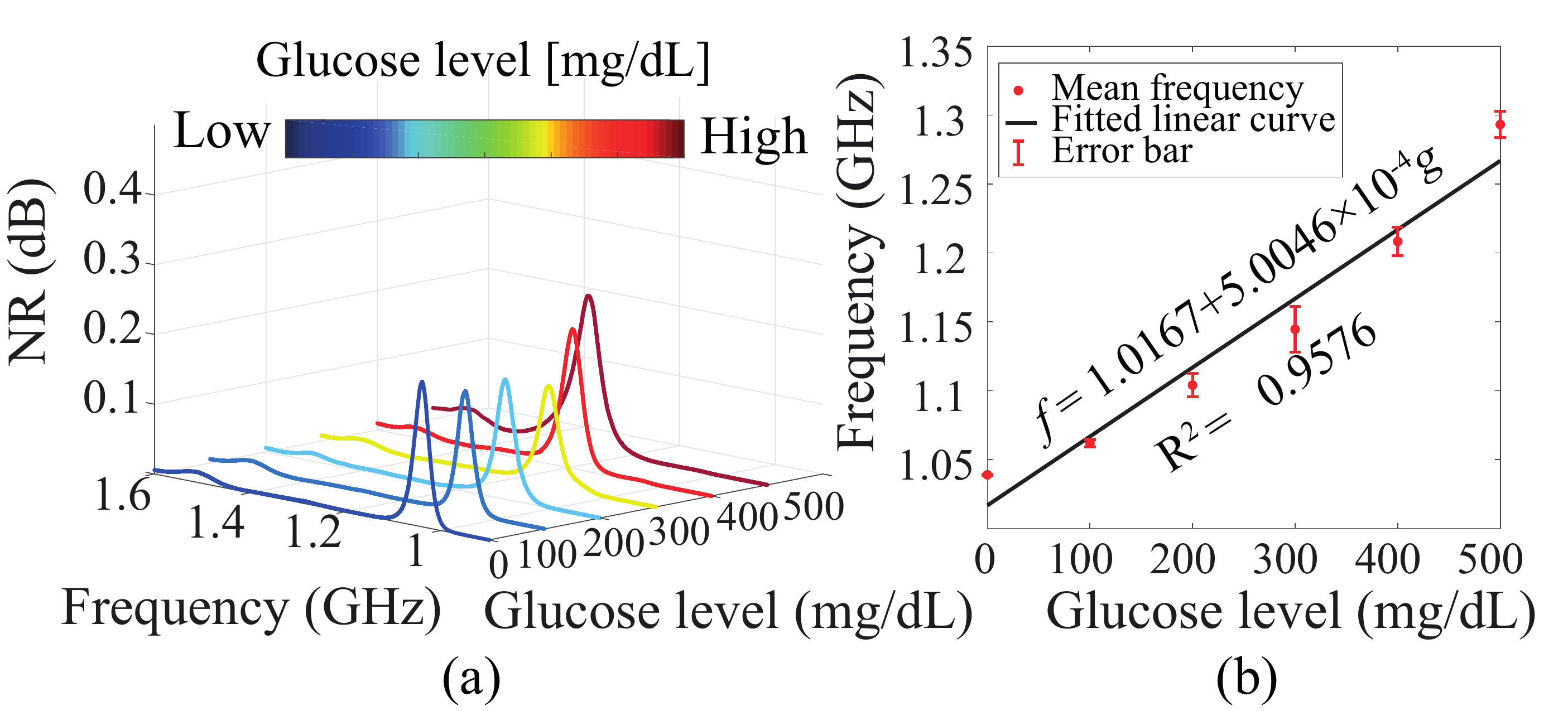}}
	\caption{ (a) Measured  sensor resonance for different glucose levels (sensor 1).  Proposed  sensor immersed in different glycaemic fluid levels.  (b) Estimated linear values of resonant frequencies for different glucose levels.  }\label{glucose}
\end{figure}

\begin{figure}[!t]
	\centering{\includegraphics[width=90mm]{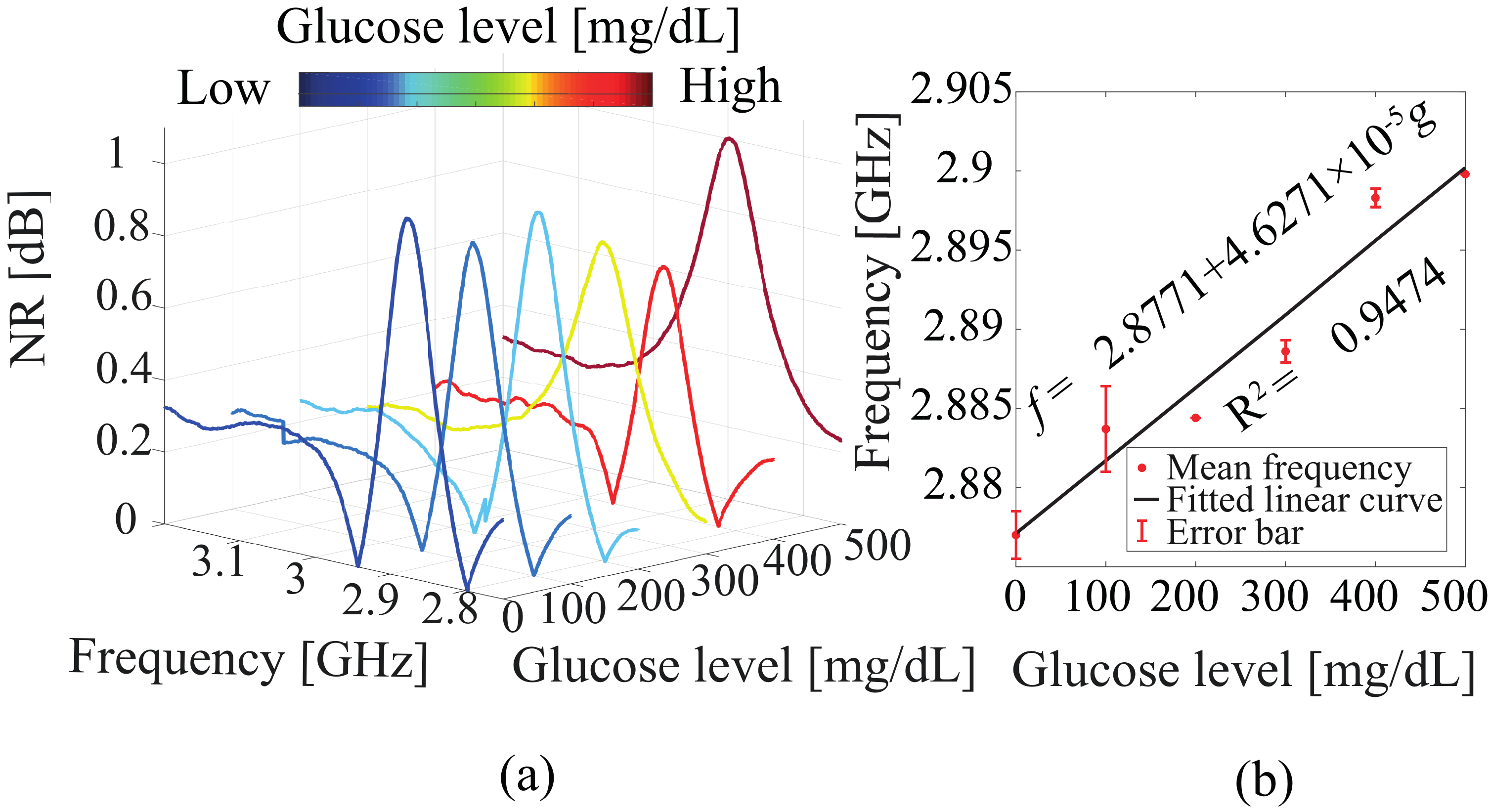}}
	\caption{(a) Measured  sensor resonance for different glucose levels (sensor 1) and different glycaemic fluid levels dropped on the top of the passive  sensor  (b) Estimated linear values of resonant frequencies for different glucose levels.} \label{glucose2}
\end{figure}

Different internal pressures  always correlate to  change in the separation  distance of the conductive  plates of the  capacitor, regardless of  using wired  \cite{pignanelli2019comparative, zhuo2017high} or wireless sensors  \cite{collins1967miniature,dehennis2002double}. Pressure sensing can also
	be used for wound monitoring. The passive  sensor  is placed on the surface of the wound and monitors mechanical pressure on the wound position \cite{agha2011review}.  In a wireless, battery-free resonator-based biosensor, the resonant frequency varies as the inverse square root of the capacitance between the conductive plates varies inversely with their separation. Thus, the frequency changes when  pressure  is introduced. An effective capacitance between the conductive plates gives rise to  any changes in the separation caused by the pressure. Thus,  we studied the effect of change in separation distance of the sensor's parallel plates on its resonance frequency.  Wound monitoring has been previously studied  by using an inductor on the top of the plate \cite{deng2018lc}. Thus, we  studied  the inductor on the top of the plate (sensor 3) in comparison to  sensors 1 and  2 in the same environment conditions (i.e., external reader, substrate).  To confirm which the passive sensor is suitable for monitoring the healing of the wound, three passive sensors with  different separations $s$ between the plates  were designed and fabricated. The separation $s$   between sensor plates varied from 0.2-0.5 mm with substrate thicknesses ranging from  0.2-0.5 mm substrate thicknesses (  i.e., top view of the passive sensor in Fig. \ref{setup}).

 $S_{11}$  were measured and the resonant frequencies of the sensors  were recorded wirelessly from the external reader  attached to the vector network analyzer (VNA)  via a coaxial cable. Responses from   $S_{11}$  were received  in the frequency range of the  resonance of the sensor, which was recorded. Each thickness was measured thrice to ensure consistency in the results. From the results, it is confirmed there is  no correlation between the separation distance of the plates,  and the change in resonance frequency of the proposed passive sensor (sensor 1), as shown in Fig. \ref{cap}(a), and proven in equation  (\ref{sensoe1_Ein}). Furthermore, the simulation results are in good agreement with measured  results.  However, in sensor 2 and sensor 3, the resonance frequency decreases with  decrease in  $s$, as shown in Fig. \ref{cap}(b) and \ref{cap}(c). The shift in the frequency $f$ is directly proportional to the decrease in the separation $s$ of the passive sensor, by an excellent correlation coefficient,   $R^2 = 0.9989$ and $R^2 = 0.99779$, for sensors 2 and 3, respectively. Moreover,  sensor 3 has a sensitivity of 99 MHz/mm, and sensor 2 has a sensitivity of 2.5$\times 10^3$ MHz/mm, which is  25 times higher than that of sensor 3. Thus, sensor 2 is most reliable for monitoring wounds as well as various internal pressures. The positive relationship between  resonance shift and  $s$  with high sensitivity is caused by an effective inner capacitor, where the electric field is dominate inside sensor 2 as proven  in equations  (\ref{sensoe2_E}), (\ref{sensoe2_C}),  and (\ref{sensoe2_f}). A slight increase  in the error bar on conducting  three different experiments of each sensor indicates that the results  can be reproduced.  To determine the advantages of the sensor 2, Table \ref{tablepressure} shows a comparison between the three passive sensors for  wound monitoring.

\begin{table}[t]
		\centering

	\caption{ Comparison between the three passive sensors for  wound sensing}
	\centering
	\vspace{5pt}
	\begin{tabular}{p{1cm}  p{1.5cm} p{2.5cm}  p{1cm}} 
		\hline \hline
		Passive sensor &	Description & Linear equation $f$ in (MHz), $s$ in (mm)& Sensitivity (MHz/mm) \\ [0.5ex] 
		\hline
		Sensor 1 & Two resonators identically charged  &\centering -  &~~~~~- \\[1ex]
		Sensor 2  & Two resonators oppositely charged     & $f = 2.107+2.5951s$ & 2.5$\times 10^3$ \\[1ex]
		Sensor 3  & Inductor on top plate    & $f = 4.07+0.0997s$& 99 \\[1ex]
		\hline\hline
	\end{tabular}
	\label{tablepressure}
\end{table}

Because of the absorption issue of the EM field on the human tissues,  especially for high frequencies and confirming the possibility for sensor 2 for implanted applications for various pressure application,  a new measurement technique using sensor 2 was prepared, wherein the sensor 2 was injected inside the phantom muscle tissue at a depth of 2~mm. The sensor was covered by an insulator material to prevent it from short-circuits when it interacting with the tissue.  The external reader was placed at a  distance of 1~mm from the tissue and 3 mm from the sensor.  A tissue sample  was then prepared, as given in \cite{yilmaz2014broadband}.
In  wireless measurements, the external reader introduces noise, as shown in Fig.   \ref{setup2}. Thus, the noise was subtracted from the  external reader signal by using the  derived noise rejection (NR) equation, which is defined  as:
\begin{equation}
NR(f) [dB]= \mid (S_{11}(f))^4_{\textit{ro}}-   (S_{11}(f))^4_{\textit{rws}} \mid
\end{equation}
\begin{table*}[t]
	\centering	

	\caption{Comparison between RF and microwave resonator-based glucose monitoring }
	\centering
	\renewcommand{\arraystretch}{2}
	\begin{tabular}{p{1.5cm}  p{1.7cm} p{1.5cm} p{2cm} p{1.5cm} p{1.5cm} p{1.5cm} p{1.5cm}} 
		\hline\hline
		[Ref.], Year & Resonator type	&  Sensing method & Structure volume (mm$^3$)&   Sample volume ($\mu$L) & Concentration (mg/dL)& Type of application for diabetic&Sensitivity (MHz/(mg/dL))\\ [0.5ex] 
		\hline
	\cite{pandit2021towards}, 2021& Localized spoof surface plasmon& Wired  &14$\times$5.03$\times$1.52 (Partial volume) &100& 0-50000
	& Outside the glycemic range  & 1.2771$\times$10$^{-6}$\\[1ex]
		\cite{kandwal2020highly}, 2020& Split
		ring & Wired  & 9.6$\times$6$\times$0.5 &~~-& 50-400& Non-invasive&3.5 \\[1ex]
		\cite{omer2020low}, 2020 & Split
		ring & Wired  &66$\times$20$\times$0.8&600& 70-110& Non-invasive&0.94 \\[1ex]
		\cite{omer2020non}, 2020 & Split
		ring & Wired  &66$\times$30$\times$0.8 &100-600& 80-120& Non-invasive & ~~-\\[1ex]
		\cite{govind2020design}, 2020 & Complementary electric LC &Wired  &80$\times$100$\times$3.175 &~95&100-500 & Non-invasive &0.0185 \\[1ex]
		\cite{kiani2021dual}, 2021 & Coplanar waveguide &Wired  &30$\times$18$\times$0.508 &~~-& 100-500& Non-invasive & 3.53-3.58\\[1ex]	
		\cite{hassan2019minimally}, 2019 &Inductor~on top plate & Wireless  & 4$\times$4$\times$1 &~~-& 0-500&  Invasive& 0.015  \\[1ex]
		\cite{hassan2020vitro}, 2020 &Cavity & Wireless & 4$\times$4$\times$2 &4000& 75-250&  Invasive& 0.032  \\[1ex]
		Proposed sensor &  Parallel resonators & Wireless  & 2$\times$ 2$\times$0.2 &30-60& 0-500& Non-invasive and invasive& 0.046-0.5 \\[1ex]
		\hline\hline
	\end{tabular}
	\label{table3}
\end{table*}where all the common noise decreased by subtracting two signals; the external  reader  (ro) signal and the reader with the sensor (rws) signal. The repeated signals that came from the reader  were neglected. As seen from Fig. \ref{signals}(a) and (b), the shift of  frequency is  directly proportional to the change in separation  with a linear regression equation $f = 1.7504+1.2762s$. Moreover, the equation comprises a statistical  R$^2$ = 0.9977, with a sensitivity of 1.3 $\times$ 10$^3$ MHz/mm.  As expected, the sensitivity was slightly reduced when the sensor was injected inside the muscle tissue. However, we can conclude that sensor 2 is sensitive to any change in the separation between the plates,  and has a significantly large detection limit because of its high capacitance with a linear detection range. Therefore,  it can be used for wound monitoring and  wide-range of pressure.   Moreover, sensor 2 is found to be well applicable inside the tissues and robust to  high capacitance with or without introducing the tissue. With the help of MEMS technologies, the FR4 in  sensor 2 can be replaced with   new,  flexible advanced materials. The MEMS passive sensors usually have lower sensitivity because of the high Young’s silicon modulus \cite{wang2017novel}. So, by introducing MEMS technologies to  sensor 2, the sensitivity is expected to reduce minimally.



\subsection{ Permittivity  monitoring (glucose monitoring)}

Glucose concentration level plays an important role in diabetes, and hence,  change in glucose levels was also studied.  A commercial electrochemical technique known as  finger-pricking devices, is commonly used to detect the glucose levels in the blood. However, in this method, the patient’s  finger is pricked   many times in the day, and hence, it can be painful for the patient.  Therefore, we require  new technologies that are  non-invasive or can be used as  long-term implantable devices, such as the proposed miniaturized  passive sensor  that  is suitable for non-invasive and minimally invasive applications, and interacts with the interstitial fluid to monitor the glucose levels.  The outside capacitance  interacts with the glycaemic fluid and creates an effective capacitance.  In previous studies,  aqueous glucose solutions were used for the initial experiments to  monitor different glucose concentrations using microwave sensors   because the  concentration of the glucose  is the most dominant as compared to the  other components of the blood \cite{kandwal2020highly, omer2020low, omer2020non, govind2020design, kiani2021dual}. Further, the glucose level in the interstitial fluid responses to changes in blood glucose \cite{jernelv2019review, basu2013time,  villena2019progress, bruen2017glucose, yilmaz2014broadband}. Thus, the glucose concentrations in aqueous solutions were prepared within the range of 0-500 mg/dL.  500 mg/dL of the stock solution was  first prepared  by mixing  glucose powder and deionized water whereas the  other concentrations were made using  deionized water and stock solution.

 The proposed  passive sensor  was immersed in a glycaemic fluid. Using a micropipette,  60 $\mu$L  glycaemic fluid was dropped into the container. The glucose concentration was  wirelessly recorded using an external reader from  a distance of 3~mm  from the passive sensor. Fig. \ref{glucose}(a) shows the responses  $S_{11}$ of the sensor under  different  concentrations of glucose. It was noticed that  the glucose level in the solution  increases, the resonance frequency also increases. The experiment was  repeated three times on different days. Further, the linear correlation between the resonant frequencies $f$ (in GHz) of the sensor and  the glucose levels $g$ (in mg/dL) is shown in Fig. \ref{glucose}(b), and the linear  regression defined  as $f=1.0167+5\times10^{-4} $. The expansion of the errors bar represent the errors noted in  the three different measurements of  glucose levels. Moreover, the equation comprises  a statistical $R^2$ = 0.9576,  with a sensitivity of  500  kHz(mg/dL) considering  different glucose levels in the aqueous solution.
 
The proposed passive sensor (sensor 1)  was further studied for use in  wearable applications in the future. The passive sensor was covered with a thick plastic tape to avoid direct contact between  the solution and  sensor. The distance between the  sensor and external reader was approximately 3 mm. The glycaemic fluid  with a volume of 30 $\mu$L  was dropped on the top of the sensor.   Further, additional different glycaemic solutions were dropped on the top of the wrapping that covered the passive  sensor. This wrapping served as a model for the skin barrier. The field from one side of the sensor could penetrate the thick tape to  interact with the glycaemic fluid.  The experiment was  repeated three times. An increase in frequency was observed, proving that as the concentration of glucose increases, the frequency increases with a sensitivity of  46  kHz(mg/dL), as shown in  Fig. \ref{glucose2}(a) and (b). Thus, we were able to demonstrate that the glucose level of the fluid can be observed by invasive or non-invasive techniques using the proposed  passive sensor.


Table \ref{table3} shows a comparison between different types of  resonators. The proposed   sensor was further  scaled down to unprecedented dimensions of 2 mm $\times$ 2 mm  $\times$ 0.2 mm to achieve the title of smallest biosensor reported for glucose monitoring,   this requiring     the smallest amount of tissue sample  to detect the glucose levels. Further, introducing a wireless,  battery-free resonator-based  glucose monitoring provides extra advantages than other  recently  reported works \cite{kandwal2020highly, omer2020low, omer2020non, govind2020design, kiani2021dual}. Further,  it has better sensitivity than the previous wired sensors, \cite{pandit2021towards, govind2020design}.  The proposed  sensor has higher sensitivity at most   30 times and 15 times more than previous wireless sensors, \cite{hassan2019minimally,hassan2020vitro}, respectively. It is compact in  size, light weight, which gives it a possibility to be applicable in  minimally invasive or non-invasive techniques, as compared to other resonators in Tabel \ref{table3}. 
 The proposed passive sensor is a promising candidate for implantable or wearable applications that monitor  glucose concentrations in the interstitial fluid.  Wireless, battery-free, and sensor-based glucose monitoring is a fundamental step for the  research  underway, and there is scope  for  commercialization  and presenting  more suitable and accurate glucose monitoring.   For example,  we can calibrate the measurement of the proposed passive sensor  for non-invasive technique and invasive technique by developing a library of the measured frequency response curves under a large variety of glucose concentrations when the passive sensor is implanted underneath the skin or attached to the skin with help of  volunteers.

\section{Conclusion \& FUTURE SCOPE}
By simply changing the position of the inductor of the proposed passive sensor based on two parallel resonators, we can change the coupling between the passive components of the sensor entirely, as well as the region of the dominance of the electric field.
Therefore, by knowing the region of the dominance of the electric field the passive sensor is no longer limited to pressure  sensing, which proves that using two parallel resonators is more effective than using an inductor on the top of the plate, where the electric field is not dominant. Three miniaturized passive sensors based on the parallel-plate technique were fabricated  to wirelessly monitor  the  wide-range biosignals, where the effective capacitive of each passive sensor was studied. The two symmetric inductors of the passive LC tank sensor with opposite direction current flow (sensor 2) created a high capacitance between the inductors, which was suitable for wound monitoring. Whereas same direction current flow (sensor 1) created a capacitance outside the inductors, making the sensor sensitive to the permittivity change of biological tissues, such as  glycaemic fluid. 

 The miniaturized passive sensors based on parallel resonators can be used for minimally invasive and non-invasive techniques, and can be  easily injected under the skin  using a rigid substrate for permittivity monitoring, i.e., sensor 1, or a flexible substrate for pressure monitoring, i.e., sensor 2.  It can also be used as a conformal structure wrapped around a specific region of the body without introducing  other elements to improve its  capacitance. The passive sensors based on parallel resonators are promising for continuous,  convenient, and faithful biosignals monitoring and an excellent alternative for all wired active resonator-based biosensors in wearable and implantable applications. Further reader development and introducing the concept of PT-symmetric and EPs to the system will improve the sensing capability of passive sensors and open the door for a  new generation of biosensors.




\ifCLASSOPTIONcaptionsoff
  \newpage
\fi




\bibliographystyle{IEEEtran}
\bibliography{sensor}
\end{document}